\documentclass[11pt]{article}

\usepackage[T1]{fontenc}
\usepackage[utf8]{inputenc}
\usepackage[margin=1in]{geometry}
\usepackage{cite}
\usepackage{amsmath,amssymb,amsfonts}
\usepackage[ruled,vlined,linesnumbered]{algorithm2e}
\usepackage{graphicx}
\usepackage{textcomp}
\usepackage{xcolor}
\usepackage{comment}
\usepackage{booktabs}
\usepackage[most]{tcolorbox}
\usepackage{url}
\usepackage{array}
\usepackage[hidelinks]{hyperref}

\providecommand{\keywords}[1]{%
  \par\vspace{0.75em}
  \noindent\textbf{Keywords:} #1
}

\begin{document}

\title{Incremental Retrieval-Augmented Generation for Human-in-the-Loop Decision Support}

\author{
Federico Mazzoni\\
Independent Researcher\\
\texttt{mazzoni.federico1@gmail.com}
\and
Andrea Failla\\
Institute of Information Science and Technologies (ISTI),\\
National Research Council (CNR), Italy\\
\texttt{andrea.failla@isti.cnr.it}
}

\date{}

\maketitle

\begin{abstract}
We introduce iRAG, a human-in-the-loop decision-support system combining Retrieval-Augmented Generation (RAG) with Hybrid Decision-Making. Unlike conventional RAG systems, relying on static Knowledge Bases (KBs), our proposal addresses settings where the KB is initially empty and is progressively populated through human decisions. The system stores previous decisions as retrievable precedents and moves from silent observation to evidence-grounded contestation and, finally, to autonomous handling with abstention. Against an independent baseline, iRAG matches final-decision error under 10\% drift without retrieval decay and reduces it under 40\% drift across all configurations. Retrieval decay improves performance on changed knowledge while increasing error on stable knowledge, evidencing a stability--plasticity trade-off rather than a general advantage. Human oversight is an architectural provision of the workflow; its usability with real human users remains to be evaluated.
\end{abstract}


\section{Introduction}
Large Language Models (LLMs) are 
trained on large text corpora to generate contextually appropriate natural language. When embedded in a larger software system to invoke tools, maintain a state, and/or act on behalf of a user, an LLM constitutes the core of an \textit{agent}~\cite{wang2024beyond}. The behaviour of an LLM-based agent depends on its training data and, at inference time, on the \textit{context} it receives: the user's input, any externally retrieved information, and a system prompt defining the agent's role and constraints~\cite{zamfirescu2023johnny}. Therefore, a frozen LLM can be adapted to different tasks without retraining by defining its context appropriately.

Retrieval-Augmented Generation (RAG) exploits this property systematically. Rather than relying on knowledge encoded during training, an LLM agent retrieves relevant passages from an external Knowledge Base (KB) at query time, and includes them in its context to produce a suitable response~\cite{gao2023retrieval}, grounding LLM outputs in verifiable, domain-specific evidence. However, standard RAG formulations assume a \textit{static} KB, compiled offline before deployment~\cite{lewis2020retrieval}, but the knowledge could be built incrementally through the decisions, progressively populating an initially empty KB through human deliberations.

The coordination of human and machine decisions over time has been extensively studied in the Machine Learning literature under the umbrella of \textit{Hybrid Decision-Making} (HDM)~\cite{punzi2024ai}, with two main paradigms: either the machine is the primary decision-maker, deferring to the human when necessary (like \textit{Learning-to-Defer, LtD})~\cite{joshi2021learning}, or the human makes decisions, and they are assisted by a model which has learned from their past behaviour (like \textit{Skeptical Learning}, SL)~\cite{zhang2019personal,zeni2019fixing}.
\textit{Bridget}~\cite{pellungrini2025bridging} unifies the two paradigms by moving between Human-in-Command and Machine-in-Command phases. The transitions are governed by \textit{Fading Empirical Accuracy} (FEA), a temporally weighted variant of the Empirical Accuracy from SL, i.e., the \textit{agreement} between the human annotator and the machine. By discounting older decisions, FEA provides an estimate of the model reliability. The original paper only proposes this idea; to the best of our knowledge, neither FEA nor the phase transition has been tested empirically. Moreover, HDM systems have dealt with \textit{concept drift}, which can also affect a RAG system built around an incremental, dynamic KB. When decision criteria evolve, older documents may misrepresent the current criteria, silently degrading the quality of RAG-grounded generations.
A static formulation of RAG is therefore ill-suited to domains where the KB must be built incrementally, and where the \textit{currency} of retrieved evidence determines decision quality.

In this work, we propose iRAG, an LLM-based system bridging RAG and HDM to support human experts when the KB must be built from their own deliberations. The system is named after both its \textit{incremental} nature, and the close relationship it has with its user(s). Our proposal rests on two main ideas.

First, \textit{the KB is not a pre-compiled corpus, but an incremental repository of previous human decisions}, structured to retrieve semantically relevant precedents for similar cases. Second, \textit{the relationship between the system and the user evolves over time}. At first, unless iRAG abstains, its decisions are compared with the human ones to update the system's FEA. Once the FEA exceeds a threshold, the system moves from a \textit{silent observer} to an \textit{active interlocutor}, presenting disagreements to the user (who retains the final decision) with evidence-grounded explanations retrieved from the KB. If FEA further improves, iRAG can move from a SL-based paradigm, in which it challenges the human, to a LtD-inspired one, in which it takes autonomous decisions and consults the human only when uncertain. Effectively, iRAG empirically tests the concept proposed by \cite{pellungrini2025bridging}.

The contributions of this paper are as follows:
\begin{itemize}
    \item We formalise \textit{incremental deliberative RAG}, i.e., RAG over a KB that grows as a direct consequence of the decisions supported by the system.
    \item We define the system behaviour, from silent observation to SL-based contestation and, finally, to LtD-inspired autonomy.
    \item We evaluate iRAG against an independent Controller-free RAG-with-defer rollout.
\end{itemize}

Our claims are deliberately system-level: we study iRAG under controlled decision-maker and acceptance policies without real human involvement, following previous controlled evaluations with simulated decision-makers or oracle-mediated in the SL
\cite{zeni2019fixing,bontempelli2020learning,Mazzoni2024frank} and LtD
\cite{mozannar2020consistent,verma2023learning,tailor2024learning}
literatures.

The rest of the paper is organised as follows. 
Section~\ref{sec:related} surveys related work across RAG and HDM systems. 
Section~\ref{sec:model} formalises the problem setting and the proposed architecture. 
Section~\ref{sec:eval} presents the experimental evaluation. 
Section~\ref{sec:conclusion} concludes with a discussion of limitations and future directions.

\section{Related Work}
\label{sec:related}
In this section, we review the literature on Hybrid Decision-Making (HDM) and Retrieval-Augmented Generation (RAG). Related topics are also discussed.

\textbf{HDM Systems.} 
Hybrid Decision-Making refers to systems where final decisions result, entirely or in part, from the collaboration between a human decision-maker and a Machine Learning model~\cite{wang2021hybrid}. HDM is especially relevant in high-stakes domains where full automation may be neither safe nor desirable. The literature identifies two broad families, corresponding to \textit{Learning-to-Abstain} and \textit{Learning Together}: \textit{machine-driven} HDM, where the model makes most decisions and defers to the human under specific conditions, and \textit{human-driven} HDM, where the human retains the primary role and the machine proposes corrections or improvements~\cite{punzi2024ai}.

The representative machine-driven paradigm is \textit{Learning-to-Defer} (LtD)~\cite{madras2018predict,mozannar2020consistent,liu2022incorporating}, an extension of \textit{Learning-to-Reject}~\cite{cortes2016learning,zhang2023survey}. Instead of simply abstaining when uncertain, the model routes ambiguous instances to a human supervisor. A limitation of machine-driven HDM is that the human is consulted at the machine's discretion and consequently plays a largely passive role.~\cite{banerjee2024learning} addresses this by providing natural-language suggestions from an integrated LLM to guide the human towards a predicted answer while preserving their final say.

Skeptical Learning (SL) belongs to the human-driven paradigm~\cite{zhang2019personal,zeni2019fixing,bontempelli2020learning,teso2021interactive,Mazzoni2024frank}. The human is the primary decision-maker, while the model learns incrementally from their decisions and acts as a \textit{surrogate}.
When its prediction differs from the human decision, it computes a \textit{skeptical score}, combining epistemic uncertainty and prediction confidence.
If it exceeds a threshold, the model presents its prediction as an alternative that the human can accept or reject; the surrogate is then updated using the final human decision. In this way, it can identify inconsistent or erroneous decisions and help the human remain consistent over time. All SL variants are affected by \textit{early-decision bias}: the surrogate is influenced by early decisions and assumes that the user acts in good faith. SL implementations differ in how the surrogate itself is updated. \cite{zeni2019fixing} retrains the model from scratch after each interaction, whereas \cite{bontempelli2020learning,Mazzoni2024frank} employ \textit{Incremental Learning} (IL) \cite{wang2023comprehensive} which updates the model with new samples rather than retraining it.
Systems that learn incrementally deal with \textit{concept drift}: a change in the underlying data distribution over time~\cite{lu2018learning,li2020incremental}.
As such, while exposed to new data, IL models need to maintain a good \textit{stability-plasticity trade-off}, where \textit{stability} refers to the ability to retain previous knowledge while learning something new, and the \textit{plasticity} to modify existing information in light of new data~\cite{kim2023stability,liang2023adaptive}.


\textbf{RAG Architecture.} Retrieval-Augmented Generation (RAG) was introduced by~\cite{lewis2020retrieval} as a general architecture for knowledge-intensive NLP tasks. Instead of relying only on knowledge encoded in the parameters of a LLM model, a RAG system employs embedding similarity to retrieve passages from an external KB at inference time and includes them in the generation context. This allows the KB to be updated without retraining the model and grounds its outputs in verifiable evidence~\cite{gao2023retrieval}. The original formulation treats the KB as static: the document index is built offline and remains fixed during training and inference~\cite{lewis2020retrieval}. Subsequent work has attempted to improve the retrieval mechanism, e.g., with \textit{reranking} techniques re-ordering retrieved documents,
thus improving the relevance of the passages passed to the main LLM \cite{asai2024self,jiang2023active,yan2024corrective}. For example, \textit{RePAQ} \cite{lewis2021paq} retrieves from a large QA-pair memory by matching the embedded input question against stored question embeddings and returning the answer associated with the nearest question. TG-RAG~\cite{han2025rag} deals with \textit{updating} the KB, by introducing temporal graph structures to handle time-sensitive knowledge, enabling incremental updates as new documents arrive.

These approaches share the assumption that the KB grows by ingesting \textit{external} document streams, such as updated domain corpora. iRAG departs from this assumption: its KB grows from \textit{human deliberations}, represented as structured records of decisions made over time. 

\section{System Overview}
\label{sec:model}
\begin{figure*}[t]
    \centering
    \includegraphics[width=\textwidth]{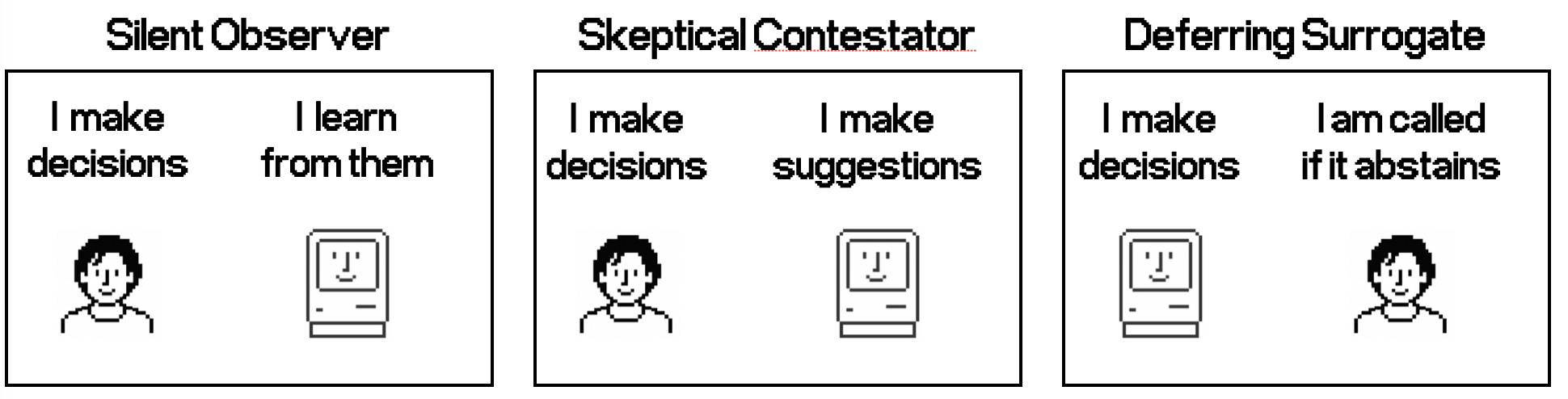}
    \caption{iRAG's three states. Transition between them is governed by
    \textit{Fading Empirical Accuracy}.}
        \label{fig:states}
\end{figure*}

We indicate with $H$ and $M$ the human user and the LLM employed by the system, and with $t_i \in T$ the $i$-th input document from a corpus $T$, e.g., a support ticket. The system also employs an auxiliary model $M_A$, which assesses whether a candidate decision \emph{covers} a reference decision. $M_A^{\mathrm{cov}}(t_C,t_R)$ is true if the content of $t_R$ is explicitly stated in or entailed by $t_C$ without contradicting it or adding additional claims (wording and structure may, however, differ).

Given $t_i$, $H$ produces the decision document $t_{H,i}$, while $M$ produces $t_{M,i}$. If the available evidence is insufficient, $M$ abstains by returning an empty $t_{M,i}$. The final decision $t_{F,i}$ corresponds to either $t_{H,i}$ or, during the later stages of the system operation, $t_{M,i}$. After each interaction, the system stores the input and its final decision directly as a question--answer record $r_i=(t_i,t_{F,i})$. The set of records forms the KB $D$, whose current size is denoted by $m=|D|$:
\begin{equation}
D=\{r_j=(t_j,t_{F,j})\}_{j=1}^{m}.
\end{equation}
Retrieval is based on the input component of each record (e.g., the question, as in \cite{lewis2021paq}): the stored input $t_j$ is embedded through an embedding model $E$:
\begin{equation}
D_E=\{E(t_j)\}_{j=1}^{m},
\end{equation}
and each record is associated with a temporal index corresponding to its insertion order: $r_1$ is the earliest record and $r_m$ the most recent. Since one question--answer record is stored for every input, $m$ increases by one at each interaction, regardless of the system state.

We distinguish $m$ from $v$, the number of reliability observations collected by the system. An observation is normally recorded whenever $M$ does not abstain and its decision can therefore be compared with the final human decision. A new negative observation is also recorded when the human flags an autonomous decision as incorrect. Consequently, $m$ and $v$ may evolve at different rates.

To generate $t_{M,i}$, iRAG first computes $e_i=E(t_i)$ and its cosine similarity with each stored-input embedding $E(t_j)\in D_E$:
\begin{equation}
c_{i,j}=\cos\!\big(e_i,E(t_j)\big)\in[-1,1].
\end{equation}
We use the native positive cosine magnitude as the semantic relevance score, thus excluding negatively correlated records:
\begin{equation}
s^+_{i,j}=\max(0,c_{i,j})\in[0,1].
\end{equation}

We then weight $s^+_{i,j}$ by an exponential decay term based on the number of records inserted into the KB after $r_j$:
\begin{equation}
\label{eq:score}
S_{i,j}
=
s^+_{i,j}\cdot\lambda_{\mathrm{RAG}}^{\,(m_i-j)},
\end{equation}
where $m_i$ is the size of the KB when $t_i$ is processed, and $\lambda_{\mathrm{RAG}}\in(0,1]$ controls how quickly the retrieval relevance of older question-answer records is discounted. The decay depends on the number of records added after $r_j$, rather than on elapsed real time.
To avoid retrieving recent but unrelated records, we require a positive raw cosine similarity satisfying $c_{i,j}>0$ and $c_{i,j}\geq\theta_{\mathrm{sem}}$. Ranking by $S_{i,j}$, including the application of temporal decay, is performed only among records satisfying this constraint.
By construction, $S_{i,j}\to0$ as $m_i-j\to\infty$ whenever $\lambda_{\mathrm{RAG}}<1$, making older records progressively harder to retrieve. With this \textit{soft temporal forgetting}, similar older records remain available in the KB (e.g., for audit), and may still be retrieved if the only relevant source of information. 
iRAG thus retrieves the top-$K$ records according to $S_{i,j}$
and includes their question and final-answer components in the context of $M$.

iRAG can operate in three states, depicted in Figure \ref{fig:states}: \textit{Silent Observer} (SO), \textit{Skeptical Contestator} (SC), and \textit{Deferring Surrogate} (DS). Initially, $D=D_E=\emptyset$ and the system is in the SO state. The KB, its embeddings, the reliability estimate, and the current state persist across interactions.
The system tracks the \textit{Fading Empirical Accuracy} (FEA) of $M$,
an HDM metric introduced in~\cite{pellungrini2025bridging}, 
here defined as
\begin{equation}
\mathrm{FEA}_v(M)
=
\frac{
    \sum_{k=1}^{v}\lambda_{\mathrm{FEA}}^{\,v-k}\delta_k
}{
    \sum_{k=1}^{v}\lambda_{\mathrm{FEA}}^{\,v-k}
}.
\end{equation}
Following SL literature, FEA measures agreement with final human decisions rather than objective task accuracy; treating it as evidence for autonomy assumes that final human decisions are sufficiently reliable. The coefficient $\lambda_{\mathrm{FEA}}\in(0,1]$ is independent of $\lambda_{\mathrm{RAG}}$: the former advances only when a reliability observation is recorded, whereas the latter advances at every KB insertion and periods without human validation do not alter FEA, which fades across observed evidence rather than elapsed interactions.
Here, $v$ is the number of reliability observations collected so far and $\delta_k\in\{0,1\}$ is the value of the $k$-th observation. When a model decision is validated against the final human decision:
\begin{equation}
\delta_k
=
\mathbb{1}
\left[
M_A^{\mathrm{cov}}(t_M,t_F)
\right].
\end{equation}
When the human flags an autonomous decision as incorrect, a new observation with $\delta_k=0$ is recorded. Following SL~\cite{zeni2019fixing,pellungrini2025bridging,Mazzoni2024frank}, agreement is measured against the human's final, possibly revised decision. By definition, $\mathrm{FEA}_v(M)\in[0,1]$, while smaller values of $\lambda_{\mathrm{FEA}}$ assign progressively less weight to older observations. In practice, $\mathrm{FEA}_v(M)$ is maintained via two running sums rather than recomputed from scratch at each step (with $A_0=W_0=0$):
\begin{equation}
\begin{aligned}
A_v &= \lambda_{\mathrm{FEA}} A_{v-1}+\delta_v, & \\
W_v &= \lambda_{\mathrm{FEA}} W_{v-1}+1, \\
\mathrm{FEA}_v(M) &= \frac{A_v}{W_v}
\quad \text{(if $W_v>0$, else $0$)},
\end{aligned}
\end{equation}

\textbf{Silent Observer.}
After $M$ generates $t_M$, the system checks whether it abstained. If $t_M$ is non-empty, $M_A$ compares it with $t_H$, a new observation $\delta$ is recorded, and $\mathrm{FEA}(M)$ is updated.
Throughout this state, $t_F\gets t_H$: the human decision is always taken as final, regardless of the model output, and $t_M$ is not shown to $H$. The system transitions to SC when $\mathrm{FEA}(M)>\alpha$ and $v\geq N$.

\textbf{Skeptical Contestator.}
$M$ first produces $t_M$. If it abstains, $t_F\gets t_H$ and no reliability observation is recorded. Otherwise, $M_A$ assesses whether $t_M$ covers $t_H$. $t_H$ is provisionally taken as $t_F$. If $M_A^{\mathrm{cov}}(t_M,t_H)$ holds, the system records $\delta=1$. Otherwise, $t_M$ is presented to the human as an alternative. If the human revises their decision, the resulting $t_F$ replaces its provisional value. The system then updates $\mathrm{FEA}(M)$ after recording $\delta=\mathbb{1}\left[M_A^{\mathrm{cov}}(t_M,t_F)\right]$. The system returns to SO if $\mathrm{FEA}(M)<\beta$, and may transition to DS if $\mathrm{FEA}(M)>\gamma$. We require $\beta<\alpha\leq\gamma$, preventing the system from oscillating between SO and SC after a single disagreement and ensuring that autonomy is not granted at a lower reliability threshold than contestation. Explicit human authorisation may trigger the transition to DS independently of $\mathrm{FEA}(M)$ and $N$.
\begin{table*}[t]
\caption{Comparison between Incremental-Learning-based HDM and iRAG.}
\label{tab:irag-il-comparison}
\centering
\scriptsize
\setlength{\tabcolsep}{4pt}
\renewcommand{\arraystretch}{0.9}
\begin{tabular}{
>{\raggedright\arraybackslash}p{0.14\textwidth}
>{\raggedright\arraybackslash}p{0.31\textwidth}
>{\raggedright\arraybackslash}p{0.31\textwidth}
}
\toprule
\textbf{Aspect} & \textbf{HDM} & \textbf{iRAG} \\
\midrule
Adapted object &
Predictive model &
External retrieval KB \\
Base model &
Updated incrementally as observations arrive &
LLM parameters remain fixed \\
New observation &
Updates model parameters, statistics, or structure &
Appended as an explicit input--decision record \\
Past observations &
Absorbed into the model state and need not be revisited &
Retained explicitly and remain individually retrievable \\
Adaptation to change &
Model updates, drift adaptation, or pruning &
Temporal weighting of retrieval relevance \\
Dependence on history &
Inference depends primarily on the learned model state &
Naive retrieval compares against all stored records \\
Auditability &
If original observations are retained separately &
Previous decisions remain explicit KB records \\
\bottomrule
\end{tabular}
\end{table*}
\textbf{Deferring Surrogate.}
$M$ becomes the primary decision-maker: for each $t$, $M$ produces $t_M$, and $H$ is consulted if $M$ abstains, in a \textit{Learning-to-Defer} fashion. Otherwise, $t_M$ becomes the final decision. Autonomous decisions do not update $\mathrm{FEA}(M)$, since they are not compared with a corresponding human decision. However, $H$ may review and override an autonomous decision. If the decision is flagged as incorrect, a new observation with $\delta=0$ is recorded and $\mathrm{FEA}(M)$ is updated. The system returns to SO if the updated $\mathrm{FEA}(M)$ is lower than $\beta$. Otherwise, it remains in DS unless the human explicitly retakes control. Like similar LtD systems, DS operates under selective feedback, with human labels observed only after deferral or review. As such, FEA may remain stale during extended autonomous operation. Periodic review is therefore required if FEA is used for continued authorization of autonomy.

\subsection{Comparison with HDM systems}
Algorithm~\ref{alg:iRAG} summarises the complete interaction loop of iRAG. As mentioned, the system shares its principles with established HDM approaches, particularly Skeptical Learning, Learning-to-Defer, and the framework proposed in~\cite{pellungrini2025bridging}, but implements adaptation through an evolving retrieval context. Whereas HDM approaches based on retraining or Incremental Learning update the predictive model using newly observed decisions, iRAG keeps the parameters of its LLM fixed and appends each input and its final decision to an external KB. Retrieved records then inform subsequent decisions without additional model training.

The use of natural-language decisions, rather than a limited set of decision labels defined \textit{a priori}, also changes how agreement and deferral are implemented. iRAG employs an auxiliary model to assess whether a proposed answer covers the final human decision, making its reliability estimate dependent on the quality of this semantic assessment. Abstention serves a distinct purpose, allowing the model to defer when the evidence retrieved from the KB is insufficient. These mechanisms preserve key HDM principles while introducing dependencies on retrieval quality and auxiliary-model judgements.

The distinction between iRAG and HDM systems is summarised in Table~\ref{tab:irag-il-comparison}. Both adapt sequentially to newly observed decisions, but they differ in what is updated and in how past observations remain available to the system. Note that if a HDM system is not employing IL, the model is always re-trained from scratch for each new observation (e.g., \cite{zeni2019fixing}).

\begin{algorithm}[t]
\scriptsize
\caption{iRAG: Incremental Retrieval-Augmented Generation}
\label{alg:iRAG}
\SetKwInOut{Input}{Input}
\SetKwInOut{Output}{Output}
\SetAlgoNoEnd
\Input{$T$; $\beta<\alpha\leq\gamma$; $N$;
$K,\theta_{\mathrm{sem}},\lambda_{\mathrm{RAG}}$;
$\lambda_{\mathrm{FEA}}$}
\Output{Final decisions $(t_{F,i})_{i=1}^{|T|}$}
$D,D_E\gets[\,]$; $\mathit{state}\gets SO$\;
$v,A,W,\mathrm{FEA}\gets 0$\;
\ForEach{$t_i\in T$}{
    $m\gets |D|$; $e_i\gets E(t_i)$\;
    $\mathcal J\gets
      \{j\in\{1,\ldots,m\}:
      \cos(e_i,D_E[j])>0\land
      \cos(e_i,D_E[j])\geq\theta_{\mathrm{sem}}\}$\;
    $\mathcal C\gets\operatorname{topK}_{j\in\mathcal J}
      (D[j];S_{i,j})$ using Eq.~\eqref{eq:score}\;
    $t_M\gets M(t_i,\mathcal C)$\;
    $\delta\gets\bot$\;
    \uIf{$\mathit{state}=DS$}{
        \uIf{$t_M=\emptyset$}{$t_F\gets H(t_i)$\;}
        \Else{
            $t_F\gets t_M$\;
            \If{$H$ flags $t_M$ as incorrect}{
                $t_F\gets H(t_i,t_M)$; $\delta\gets0$\;
            }
        }
    }
    \Else{
        $t_H\gets H(t_i)$; $t_F\gets t_H$\;
        \If{$t_M\neq\emptyset$}{
            $b\gets M_A^{\mathrm{cov}}(t_M,t_H)$\;
            \uIf{$\mathit{state}=SC\land\neg b$}{
                Present $(t_M,\mathcal C)$ to $H$\;
                $t_F\gets H(t_i,t_M)$\;
                $\delta\gets\mathbb{1}
                  [M_A^{\mathrm{cov}}(t_M,t_F)]$\;
            }
            \Else{$\delta\gets\mathbb{1}[b]$\;}
        }
    }
    \If{$\delta\neq\bot$}{
        $v\gets v+1$\;
        $A\gets\lambda_{\mathrm{FEA}}A+\delta$\;
        $W\gets\lambda_{\mathrm{FEA}}W+1$\;
        $\mathrm{FEA}\gets A/W$\;
    }
    \uIf{$\mathit{state}=DS\land
      (\mathit{retake}_H\lor
      (\delta=0\land\mathrm{FEA}<\beta))$}{
        $\mathit{state}\gets SO$\;
    }
    \ElseIf{$\mathit{state}\neq DS\land\mathit{authorise}_H$}{
        $\mathit{state}\gets DS$\;
    }
    \ElseIf{$\mathit{state}=SO\land v\geq N\land
      \mathrm{FEA}>\alpha$}{
        $\mathit{state}\gets SC$\;
    }
    \ElseIf{$\mathit{state}=SC\land\mathrm{FEA}<\beta$}{
        $\mathit{state}\gets SO$\;
    }
    \ElseIf{$\mathit{state}=SC\land\mathrm{FEA}>\gamma$}{
        $\mathit{state}\gets DS$\;
    }
    Append $(t_i,t_F)$ to $D$; append $e_i$ to $D_E$\;
    $t_{F,i}\gets t_F$\;
}
\end{algorithm}

\section{Evaluation}
\label{sec:eval}

iRAG was evaluated on a synthetic benchmark simulating customer-support tickets for a CRM-like product undergoing quarterly releases, with simulated human profiles.
This evaluation does not substitute for a human-participant study, which we defer to future work. The experiment separates the effects of retrieval and FEA decay on adaptation, authority allocation, and final-decision accuracy. The code with instructions to replicate the experiments is available on: \url{https://github.com/FedericoMz/iRAG}.

\textbf{Dataset.}
We created a synthetic dataset representing four quarters $Q_1,\dots,Q_4$ of product evolution. Each quarter contains $500$ tickets across five topics (\textit{Billing, Integrations, Permissions, Reporting}, and \textit{Onboarding}) and tagged as easy, normal, or hard. 
A deterministic pipeline (available for audit in the project repo) generated the product documentation, the tickets, a gold answer consistent with the documentation of the corresponding quarter, and a perturbed answer for each human profile. The first quarter contains no drift and is shared by both dataset variants. In $Q_2$--$Q_4$, either 50 (10\%) or 200 (40\%) tickets are near-duplicates of earlier tickets whose correct answer changed after a release. These tickets form the \textit{drift subset}; the others the \textit{stable subset}. The two variants contain, respectively, $150$ and $600$ drift tickets, among the $2,000$ nominal tickets.

\textbf{Human Profiles.}
We simulate three decision-makers by perturbing the gold answers,
following the methodology used in prior HDM evaluations
\cite{Mazzoni2024frank,mozannar2020consistent,tailor2024learning}: the \textit{CEO}, who is always correct; the \textit{Domain Expert} (DE), reliable within one declared topic; and the \textit{Intern}, reliable on easy tickets and systematically defaults to the previous quarter's answer on drift tickets.
Hard tickets and a randomly selected 50\% of out-of-domain normal tickets are assigned to the CEO; in-domain easy and normal tickets are given to the DE; out-of-domain easy tickets are answered by the Intern; and the remaining out-of-domain normal tickets are assigned to the DE acting outside their expertise, set as \textit{Billing}. The random assignment is resampled independently in each repetition. Following the original SL evaluation, where a related oracle-mediated resolution was used \cite{zeni2019fixing}, in SC a suggestion is accepted if it covers the gold answer (as checked by an auxiliary LLM call).
This is an experimental control policy, simulating cooperative users who accept a disputed proposal if judged as correct. 

\textbf{Experimental Protocol.}
Quarters are processed in order to preserve drift, while tickets within each quarter are shuffled. Every run begins with a trusted bootstrap: all $Q_1$ tickets are assigned to the CEO and iRAG is locked in SO while constructing the KB and updating FEA. Normal state transitions and informed-mixture assignment begin in $Q_2$. The first 100 tickets of every subsequent quarter are also assigned to the CEO, and state transitions are suspended until this review window ends. If a quarter begins in DS, iRAG first returns to SC. This periodic review provides recent trusted decisions, potentially including drift tickets because the within-quarter order is random, and mitigates the early-decision bias of SL-based systems.
We independently set $\lambda_{\mathrm{RAG}}$ and $\lambda_{\mathrm{FEA}}$ either to $0.99861$ (D) or to $1$ (ND), yielding four decay combinations at each drift level. The decaying value gives retrieval a half-life of approximately one quarter (500 KB insertions) and FEA a half-life of approximately 500 reliability observations. This factorial design separates retrieval plasticity from adaptation of the reliability estimate. We set $\theta_{\mathrm{sem}}=0.7$, $K=5$, $(\alpha,\beta,\gamma)=(0.70,0.55,0.80)$, and $N=30$. Both $M$ and $M_A$ use Amazon Nova 2 Lite through Amazon Bedrock with temperature zero, chosen for reproducibility of the decision policy; sensitivity to sampling temperature is left to future work\footnote{Although $M$ and $M_A$ share the same underlying model, their calls are stateless and use distinct prompts and output schemas. Across the experiments, $M_A$ rejected $8{,}159/16{,}064$ disputed proposals (50.79\%) and classified $30{,}102/151{,}112$ non-abstaining, gold-evaluated model answers as incorrect (19.92\%). It rejected $830/1{,}077$ answers that exactly reproduced a dataset-labelled incorrect human response (77.07\%), while accepting all $19{,}062$ exact matches to the gold answer. This audit
provides no evidence of blanket preference for $M$'s outputs.}. Questions were embedded with \texttt{qwen3-embedding:4b} through Ollama. Each configuration is repeated ten times using the same initial seed and independent within-quarter permutations.

\begin{figure*}[p]
    \centering
    \includegraphics[width=\textwidth,height=0.84\textheight,keepaspectratio]{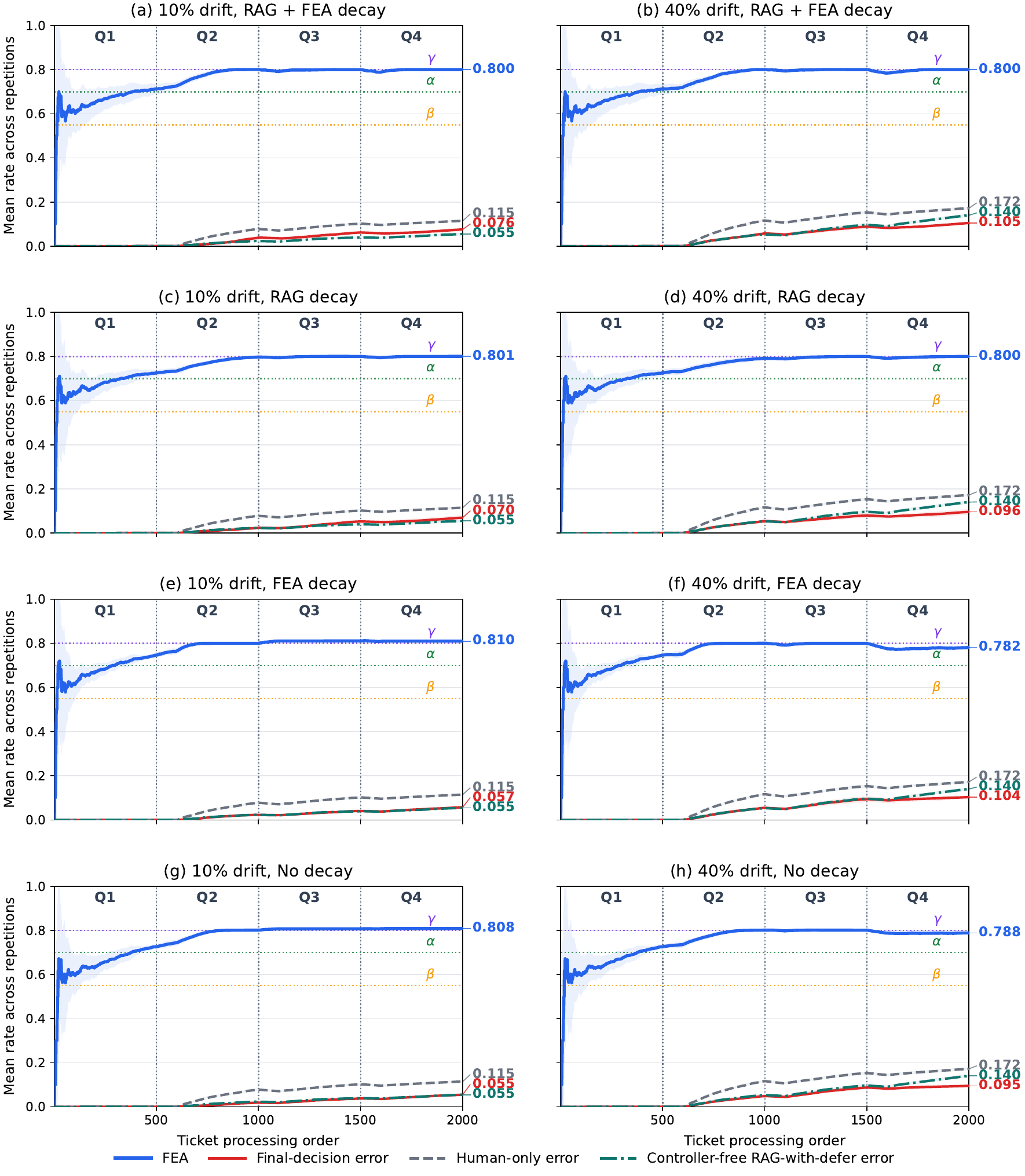}
    \caption{Mean trajectories across ten repetitions for all eight
    configurations. Rows compare coupled decay, RAG-only decay, FEA-only
    decay, and no decay; columns compare 10\% (left) and 40\% (right) drift.
    Shaded bands show one population standard deviation, vertical dotted lines
    separate quarters, and horizontal dashed lines mark the state thresholds.
    Each column repeats the corresponding independent, non-decaying Controller-free
    RAG-with-defer trajectory.}
    \label{fig:all-settings}
\end{figure*}

\textbf{Evaluation Measures.}
The \textit{final-decision error} is the proportion of final answers not covering the gold answer. The \textit{human-only baseline} is the counterfactual error obtained by always retaining the initially assigned human answer. Our \textit{Controller-free RAG-with-defer} baseline is instead executed as an independent end-to-end simulation. It follows the same bootstrap and review protocol: the CEO finalizes all $Q_1$ tickets and the first 100 tickets of each later quarter. On every remaining ticket, $M$ uses the same retrieval and abstention rules as iRAG; a non-abstaining answer is final, whereas an abstention defers to the informed-mixture-assigned human. The baseline constructs its own KB exclusively from its own final decisions, uses non-decaying retrieval ($\lambda_{\mathrm{RAG}}=1$), and has no SO/SC/DS controller or FEA. We run it with the same ten seeds; consequently, one independent baseline is shared by all four iRAG configurations within each drift dataset.
We report final-decision error separately for stable and drift tickets. The \textit{abstention rate} is computed conditionally on the tickets processed in SO, SC, or DS. The \textit{SC proposal-acceptance rate} is the number of accepted proposals divided by the number of explicit accept/reject decisions, excluding model abstentions and model answers already covering the human answer. Finally, \textit{model-finalized tickets} is the proportion of all nominal tickets for which the autonomous model supplies the final answer. 
Table~\ref{tab:experimental-results} reports the mean and sample standard deviation across repetitions. For the primary outcomes, paired 95\% Student-$t$ confidence intervals and two-sided exact sign-flip tests with Holm correction are reported below when relevant. Per-run metrics, counts, statistical tests, plots, and seeds are provided in the project repository.

\begin{table*}[t]
\caption{Mean $\pm$ sample standard deviation across ten repetitions.
Rates are percentages except for FEA; final-state entries are repetition
counts. R and F denote RAG and FEA decay, respectively; R:ND is the
incremental-RAG benchmark ($\lambda_{\mathrm{RAG}}=1$). Controller-free RAG uses
$\lambda_{\mathrm{RAG}}=1$, so its dataset-level result spans all four columns.}
\label{tab:experimental-results}
\centering
\scriptsize
\setlength{\tabcolsep}{1.5pt}
\renewcommand{\arraystretch}{0.82}
\resizebox{\textwidth}{!}{%
\begin{tabular}{@{}l*{8}{c}@{}}
\toprule
& \multicolumn{4}{c}{10\%}
& \multicolumn{4}{c}{40\%} \\
\cmidrule(lr){2-5}
\cmidrule(lr){6-9}
Metric
& \shortstack{R:D\\F:D}
& \shortstack{R:D\\F:ND}
& \shortstack{R:ND\\F:D}
& \shortstack{R:ND\\F:ND}
& \shortstack{R:D\\F:D}
& \shortstack{R:D\\F:ND}
& \shortstack{R:ND\\F:D}
& \shortstack{R:ND\\F:ND} \\
\midrule

Human-only baseline
& \multicolumn{4}{c}{$11.47\pm0.52$}
& \multicolumn{4}{c}{$17.22\pm0.68$} \\

Controller-free RAG-with-defer error
& \multicolumn{4}{c}{$5.50\pm0.62$}
& \multicolumn{4}{c}{$14.03\pm0.73$} \\

Final-decision error
& $7.59\pm0.97$ & $6.99\pm1.20$ & $5.75\pm0.42$ & $5.52\pm0.41$
& $10.52\pm0.68$ & $9.62\pm1.43$ & $10.36\pm0.87$ & $9.51\pm0.82$ \\

Abstain (SO)
& $15.73\pm0.65$ & $15.68\pm0.59$ & $14.97\pm0.36$ & $15.17\pm0.32$
& $15.71\pm0.71$ & $15.70\pm0.62$ & $14.97\pm0.36$ & $15.12\pm0.36$ \\

Abstain (SC)
& $2.63\pm0.72$ & $2.28\pm0.58$ & $0.03\pm0.10$ & $0.02\pm0.08$
& $2.83\pm0.54$ & $2.27\pm0.40$ & $0.08\pm0.09$ & $0.02\pm0.07$ \\

Abstain (DS)
& $3.36\pm0.63$ & $2.01\pm0.52$ & $0.08\pm0.15$ & $0.12\pm0.12$
& $3.01\pm0.86$ & $2.32\pm0.99$ & $0.03\pm0.07$ & $0.00\pm0.00$ \\

\shortstack[l]{Accepted proposals (SC)}
& $52.39\pm2.29$ & $56.09\pm3.61$ & $33.95\pm10.48$ & $53.73\pm7.69$
& $50.39\pm2.84$ & $53.26\pm2.97$ & $40.23\pm2.77$ & $45.55\pm2.67$ \\

Stable-ticket error
& $6.08\pm0.62$ & $5.51\pm0.97$ & $3.45\pm0.37$ & $3.20\pm0.38$
& $5.02\pm0.90$ & $3.94\pm1.36$ & $3.17\pm0.77$ & $2.65\pm0.55$ \\

Drift-ticket error
& $26.20\pm5.87$ & $25.20\pm5.65$ & $34.07\pm2.92$ & $34.07\pm2.00$
& $23.33\pm1.42$ & $22.87\pm1.87$ & $27.12\pm1.81$ & $25.50\pm1.88$ \\

\shortstack[l]{Model-finalized tickets}
& $43.34\pm4.36$ & $36.45\pm7.30$ & $55.93\pm1.61$ & $50.70\pm1.88$
& $35.80\pm4.67$ & $27.42\pm10.74$ & $32.09\pm3.50$ & $26.12\pm2.99$ \\

Final FEA
& $0.8\pm0.0001$ & $0.8\pm0.002$ & $0.81\pm0.004$ & $0.81\pm0.005$
& $0.8\pm0.00008$ & $0.8\pm0.00002$ & $0.78\pm0.01$ & $0.79\pm0.005$ \\

Final state (DS/SC)
& 10/0 & 10/0 & 10/0 & 10/0
& 10/0 & 10/0 & 1/9 & 0/10 \\

\bottomrule
\end{tabular}%
}
\end{table*}

\subsection{Contribution of RAG Decay}

Figure~\ref{fig:all-settings} reports the FEA and cumulative-error trajectories for the complete factorial design. During the CEO-controlled $Q_1$, iRAG's final-decision and human-only error remain zero by construction while FEA accumulates evidence about model behaviour. The initially wide FEA bands narrow as observations accumulate. From $Q_2$ onward, the configurations diverge under normal assignment and state transitions; the local changes around later quarter boundaries reflect the injection of recent CEO-reviewed decisions before transitions resume. The independently rolled-out Controller-free RAG-with-defer trajectory is repeated across the four panels for the corresponding drift dataset, rather than recomputed from each iRAG trajectory.

Against the principal incremental-RAG benchmark ($\lambda_{\mathrm{RAG}}=1$), retrieval decay produces the stability--plasticity trade-off. Holding FEA decay fixed, it lowers drift-ticket error by $7.87$--$8.87$ points under 10\% drift (paired 95\% CIs $[-11.10,-4.63]$ and $[-12.16,-5.57]$) and by $2.63$--$3.78$ points under 40\% drift ($[-4.00,-1.27]$ and $[-5.29,-2.28]$). All four drift-ticket improvements remain significant after Holm correction. Retrieval decay simultaneously raises stable-ticket error by $2.31$--$2.62$ and $1.29$--$1.85$ points, respectively, with all four paired CIs above zero. The mechanism is consistent with the scoring rule: down-weighting older precedents reduces interference from superseded decisions, but also weakens still-valid evidence when no change has occurred.

The aggregate result depends on how many tickets can benefit from this plasticity. The 10\% dataset contains $150$ drift and $1{,}850$ stable tickets. Per repetition, retrieval decay therefore prevents approximately $12$--$13$ drift errors but introduces approximately $43$--$49$ stable errors, increasing aggregate error by $1.48$--$1.84$ points. The 40\% dataset contains $600$ drift and $1{,}400$ stable tickets: approximately $16$--$23$ avoided drift errors are offset by $18$--$26$ additional stable errors, leaving only $0.11$--$0.16$ points of aggregate difference, with both CIs including zero. Thus, the smaller percentage-point improvement on individual drift tickets at 40\% drift has greater system-level relevance because it applies to four times as many changed tickets. The incremental-RAG benchmark retains the lowest mean aggregate error, whereas retrieval decay consistently provides stronger resistance to concept drift. Retrieval decay is therefore not universally superior, but a controllable exchange of stability for plasticity whose utility grows with the prevalence and cost of change.

\subsection{Contribution of FEA Decay}

FEA decay has a different, primarily operational effect. It does not directly alter which precedents are retrieved; instead, it reduces the influence of older agreement observations on the reliability estimate controlling state transitions. Holding retrieval decay fixed, FEA decay increases the model-finalized proportion by $5.24$--$6.89$ points under 10\% drift and $5.98$--$8.38$ points under 40\% drift, with every paired CI above zero. Over $2{,}000$ tickets, this corresponds to approximately $105$--$138$ additional model-finalized decisions per repetition under 10\% drift and $120$--$168$ under 40\% drift. FEA decay should therefore be interpreted primarily as a change in the timing and extent of delegation, rather than as a direct improvement to retrieval quality.

The accuracy consequence depends on the environment. Under 10\% drift, FEA decay changes final-decision error by only $0.24$--$0.60$ points and both CIs include zero. Under 40\% drift, it increases error by $0.85$ points in the incremental-RAG benchmark (95\% CI $[0.21,1.49]$) and by $0.90$ points with retrieval decay ($[0.19,1.61]$), although only the former remains significant after within-contrast Holm correction. This is consistent with an authority-exposure effect: earlier or longer residence in DS makes a larger share of model errors final, even though the decay operation itself does not directly change the retrieved evidence.

The final states make the threshold sensitivity visible. All 10\% repetitions finish in DS. Under 40\% drift, both retrieval-decay conditions finish in DS in all ten repetitions, compared with one of ten for FEA decay alone and none for the fully non-decaying benchmark. Yet the terminal mean FEA values occupy a narrow range around the predefined $\gamma=0.80$. Small continuous differences can therefore produce large differences in the discrete DS/SC count, and the final state alone does not reveal how long a run spent under autonomous authority. We consequently treat final-state frequencies as threshold-dependent operational outcomes and rely on continuous FEA, error, and model-finalization rates as the principal evidence. This motivates future sensitivity analysis of $\alpha$, $\beta$, and $\gamma$ rather than interpreting the observed state split as a threshold-independent property of the system.

\subsection{Overall Contribution of iRAG}

Controller-free RAG-with-defer supplies a fully independent comparison. Its non-decaying KB is updated from its own decisions, so the comparison includes the downstream effect of authority allocation on later retrieval rather than replaying iRAG's realised trajectory. It obtains $5.50\pm0.62$\% final-decision error under 10\% drift and $14.03\pm0.73$\% under 40\% drift. Because this baseline always uses $\lambda_{\mathrm{RAG}}=1$, the same dataset-level value is compared with all four iRAG configurations; only the R:ND columns hold the retrieval-decay setting fixed.

Under 10\% drift, Controller-free RAG-with-defer is numerically best. The two R:ND iRAG settings obtain $5.52$--$5.75$\%, and their paired differences from the baseline are small ($0.02$--$0.25$ points) with both 95\% CIs crossing zero. With retrieval decay, iRAG's aggregate error increases to $6.99$--$7.59$\%: its drift-ticket error is lower than the baseline ($25.20$--$26.20$\% versus $31.67$\%), but its stable-ticket error is higher ($5.51$--$6.08$\% versus $3.38$\%). Thus, at low drift, iRAG matches the simpler baseline without retrieval decay, while temporal plasticity incurs the stability cost identified above.

The comparison reverses under 40\% drift. Every iRAG configuration scores $9.51$--$10.52$\%, reducing error by $3.52$--$4.53$ points relative to Controller-free RAG-with-defer; all paired 95\% CIs have lower bounds of at least $2.92$ points, and all ten matched seeds favour iRAG in every configuration. Here iRAG improves both stable-ticket error ($2.65$--$5.02$\% versus $7.84$\%) and drift-ticket error ($22.87$--$27.12$\% versus $28.47$\%). It also allows the model to finalize only $26.12$--$35.80$\% of tickets, compared with $59.98$\% for the controller-free baseline, showing that the gain coincides with more selective autonomy rather than more model exposure.

Taken together, the results bound iRAG's contribution more precisely. The controller does not yield an unconditional advantage: under mild drift, an independent non-decaying RAG-with-defer policy is already competitive. Under stronger drift, however, iRAG's state-dependent allocation of authority produces a substantial end-to-end improvement over both that baseline and the human-only policy. The factorial design then separates the remaining mechanisms: retrieval decay governs the stability--plasticity trade-off, while FEA decay governs the speed and extent of delegation. Per-run metrics, counts, statistical tests, plots, and seeds are provided in the project repository.

\section{Conclusion}
\label{sec:conclusion}
We have presented iRAG, an incremental deliberative RAG system in which the knowledge base is constructed directly from the decisions supported by the system. iRAG operationalises a progressive collaboration between humans and LLMs: it first observes human decisions, then contests unsupported disagreements, and finally assumes conditional autonomy. In doing so, it combines an auditable, continuously updated RAG memory with explicit transitions between human-in-command and machine-in-command operation.

The experiments provide the first empirical evaluation of FEA-driven transitions. Against independently rolled-out Controller-free RAG-with-defer, iRAG is statistically indistinguishable in the mild-drift, non-decaying setting and clearly more accurate under stronger drift. This is achieved by granting the model less final authority, establishing a synergy between the system and the human user(s) (defined as a \textit{co-evolutionary relationship} in \cite{Mazzoni2024frank}). The complete factorial design separates the two forms of temporal decay: retrieval decay produces the stability--plasticity trade-off, consistently improving changed knowledge while sacrificing some stable accuracy, whereas FEA decay chiefly increases the amount of delegated authority. Under stronger drift, the retrieval benefit almost entirely offsets its stability cost and enables every retrieval-decay repetition to return to autonomous operation. iRAG therefore provides a concrete, auditable mechanism for managing adaptation and authority in an incremental RAG system while retaining human oversight, model abstention, and periodic reassessment.

By design, iRAG also inherits assumptions from the HDM systems it combines. From SL, it inherits reliance on human decisions, and therefore early-decision bias (the reason for the CEO bootstrap and periodic trusted review in our experiments), as well as an \textit{empirical accuracy} measured through \textit{agreement} rather than objective correctness. From LtD, the Deferring Surrogate state inherits the risk of over-reliance on the machine and selective feedback during autonomous operation. From \textit{Bridget}, it inherits user-specified, potentially arbitrary thresholds for state transitions, whose sensitivity is visible in the final-state results. These are explicit design assumptions inherited from the original paradigms and limit the conclusions that can be drawn from the present evaluation. Future work can mitigate them through learned or task-calibrated thresholds, broader and independent reliability checks, and adaptive human-review schedules. Moreover, as mentioned, $M$ and $M_A$ share the same underlying model family; although our audit (see fn. 1) found no evidence of blanket self-preference, a fully independent auxiliary judge would strengthen this validation further. Additionally, our experiments assign a substantial share of tickets to the oracle decision-maker; future works should vary the quantity, to establish how iRAG behaves with a fallible bootstrap expert or narrower review windows.

This study establishes the feasibility of the architecture under controlled conditions and provides a reproducible basis for its human-centred validation. As the next step, we are planning a workshop with human participants to evaluate the transition and contestation mechanisms in practice and to begin the participatory design of the iRAG interface. The workshop will examine how users understand system states, inspect retrieved evidence, respond to contested decisions, and take back or delegate control.

\bibliographystyle{abbrv}
\bibliography{main}

@article{joshi2021learning,
  author       = {Shalmali Joshi and
                  others},
  title        = {Pre-emptive learning-to-defer for sequential medical decision-making
                  under uncertainty},
  journal      = {CoRR},
  volume       = {abs/2109.06312},
  year         = {2021}
}

@inproceedings{zhang2019personal,
  author       = {Wanyi Zhang},
  title        = {Personal Context Recognition via Skeptical Learning},
  booktitle    = {{IJCAI}},
  pages        = {6482--6483},
  publisher    = {ijcai.org},
  year         = {2019}
}

@inproceedings{bontempelli2020learning,
  author       = {Andrea Bontempelli and
                  others},
  title        = {Learning in the Wild with Incremental Skeptical Gaussian Processes},
  booktitle    = {{IJCAI}},
  pages        = {2886--2892},
  publisher    = {ijcai.org},
  year         = {2020}
}

@inproceedings{teso2021interactive,
  author       = {Stefano Teso and
                  others},
  title        = {Interactive Label Cleaning with Example-based Explanations},
  booktitle    = {NeurIPS},
  pages        = {12966--12977},
  year         = {2021}
}

@article{wang2023comprehensive,
author       = {Liyuan Wang and
                  others},
  title        = {A Comprehensive Survey of Continual Learning: Theory, Method and Application},
  journal      = {CoRR},
  volume       = {abs/2302.00487},
  year         = {2023}
}

@article{zeni2019fixing,
  author       = {Mattia Zeni and
                  others},
  title        = {Fixing Mislabeling by Human Annotators Leveraging Conflict Resolution
                  and Prior Knowledge},
  journal      = {{ACM} {IMWUT}},
  volume       = {3},
  number       = {1},
  pages        = {32:1--32:23},
  year         = {2019}
}

@inproceedings{mozannar2020consistent,
  author       = {Hussein Mozannar and
                  others},
  title        = {Consistent Estimators for Learning to Defer to an Expert},
  booktitle    = {{ICML}},
  series       = {Proceedings of Machine Learning Research},
  volume       = {119},
  pages        = {7076--7087},
  publisher    = {{PMLR}},
  year         = {2020}
}

@article{punzi2024ai,
  title={AI, Meet Human: Learning Paradigms for Hybrid Decision Making Systems},
  author={Punzi, Clara and
                  others},
  journal={arXiv preprint arXiv:2402.06287},
  year={2024}
}

@article{banerjee2024learning,
  title={Learning To Guide Human Decision Makers With Vision-Language Models},
  author={Banerjee, Debodeep and
                  others},
  journal={arXiv preprint arXiv:2403.16501},
  year={2024}
}

@inproceedings{Mazzoni2024frank,
  author       = {Federico Mazzoni and
                  others},
  title        = {A Frank System for Co-Evolutionary Hybrid Decision-Making},
  booktitle    = {{IDA} {(2)}},
  series       = {Lecture Notes in Computer Science},
  volume       = {14642},
  publisher    = {Springer},
  year         = {2024}
}

@article{wang2021hybrid,
  author       = {Tong Wang and
                  others},
  title        = {Hybrid Predictive Models: When an Interpretable Model Collaborates
                  with a Black-box Model},
  journal      = {J. Mach. Learn. Res.},
  volume       = {22},
  pages        = {137:1--137:38},
  year         = {2021}
}

@article{zhang2023survey,
  author       = {Xu{-}Yao Zhang and
                  others},
  title        = {A Survey on Learning to Reject},
  journal      = {Proc. {IEEE}},
  volume       = {111},
  number       = {2},
  pages        = {185--215},
  year         = {2023}
}

@inproceedings{kim2023stability,
  title={On the Stability-Plasticity Dilemma of Class-Incremental Learning},
  author={Kim, Dongwan and
                  others},
  booktitle={Proceedings of the IEEE/CVF Conference on Computer Vision and Pattern Recognition},
  pages={20196--20204},
  year={2023}
}

@inproceedings{liang2023adaptive,
  title={Adaptive Plasticity Improvement for Continual Learning},
  author={Liang, Yan-Shuo and
                  others},
  booktitle={Proceedings of the IEEE/CVF Conference on Computer Vision and Pattern Recognition},
  pages={7816--7825},
  year={2023}
}

@article{li2020incremental,
  title={Incremental learning imbalanced data streams with concept drift: The dynamic updated ensemble algorithm},
  author={Li, Zeng and
                  others},
  journal={Knowledge-Based Systems},
  volume={195},
  pages={105694},
  year={2020},
  publisher={Elsevier}
}

@article{lu2018learning,
  title={Learning under concept drift: A review},
  author={Lu, Jie and
                  others},
  journal={IEEE transactions on knowledge and data engineering},
  volume={31},
  number={12},
  pages={2346--2363},
  year={2018},
  publisher={IEEE}
}

@article{liu2022incorporating,
  title={Incorporating uncertainty in learning to defer algorithms for safe computer-aided diagnosis},
  author={Liu, Jessie and
                  others},
  journal={Scientific reports},
  volume={12},
  number={1},
  pages={1762},
  year={2022},
  publisher={Nature Publishing Group UK London}
}

@article{madras2018predict,
  title={Predict responsibly: improving fairness and accuracy by learning to defer},
  author={Madras, David and
                  others},
  journal={Advances in Neural Information Processing Systems},
  volume={31},
  year={2018}
}

@inproceedings{cortes2016learning,
  title={Learning with rejection},
  author={Cortes, Corinna and
                  others},
  booktitle={Algorithmic Learning Theory: 27th International Conference, ALT 2016, Bari, Italy, October 19-21, 2016, Proceedings 27},
  pages={67--82},
  year={2016},
  organization={Springer}
}

@inproceedings{zamfirescu2023johnny,
  title={Why Johnny can’t prompt: how non-AI experts try (and fail) to design LLM prompts},
  author={Zamfirescu-Pereira, JD and
                  others},
  booktitle={Proceedings of the 2023 CHI Conference on Human Factors in Computing Systems},
  pages={1--21},
  year={2023}
}

@article{gao2023retrieval,
  title={Retrieval-augmented generation for large language models: A survey},
  author={Gao, Yunfan and
                  others},
  journal={arXiv preprint arXiv:2312.10997},
  year={2023}
}

@article{wang2024beyond,
  title={Beyond the limits: A survey of techniques to extend the context length in large language models},
  author={Wang, Xindi and
                  others},
  journal={arXiv preprint arXiv:2402.02244},
  year={2024}
}

@article{lewis2020retrieval,
  title={Retrieval-augmented generation for knowledge-intensive nlp tasks},
  author={Lewis, Patrick and
                  others},
  journal={Advances in neural information processing systems},
  volume={33},
  pages={9459--9474},
  year={2020}
}

@incollection{pellungrini2025bridging,
  title={Bridging the Gap in Hybrid Decision-Making Systems},
  author={Pellungrini, Roberto and
                  others},
  booktitle={HHAI 2025},
  pages={41--48},
  year={2025},
  publisher={IOS Press}
}

@inproceedings{asai2024self,
  title={Self-rag: Learning to retrieve, generate, and critique through self-reflection},
  author={Asai, Akari and
                  others},
  booktitle={International conference on learning representations},
  volume={2024},
  pages={9112--9141},
  year={2024}
}

@inproceedings{jiang2023active,
  title={Active retrieval augmented generation},
  author={Jiang, Zhengbao and
                  others},
  booktitle={Proceedings of the 2023 conference on empirical methods in natural language processing},
  pages={7969--7992},
  year={2023}
}

@article{yan2024corrective,
  title={Corrective retrieval augmented generation},
  author={Yan, Shi-Qi and
                  others},
  year={2024}
}

@article{han2025rag,
  title={RAG Meets Temporal Graphs: Time-Sensitive Modeling and Retrieval for Evolving Knowledge},
  author={Han, Jiale and
                  others},
  journal={arXiv preprint arXiv:2510.13590},
  year={2025}
}

@article{lewis2021paq,
  title={PAQ: 65 million probably-asked questions and what you can do with them},
  author={Lewis, Patrick and
                  others},
  journal={Transactions of the Association for Computational Linguistics},
  volume={9},
  pages={1098--1115},
  year={2021},
  publisher={MIT Press One Rogers Street, Cambridge, MA 02142-1209, USA journals-info~…}
}

@inproceedings{verma2023learning,
  title={Learning to defer to multiple experts: Consistent surrogate losses, confidence calibration, and conformal ensembles},
  author={Verma, Rajeev and
                  others},
  booktitle={International Conference on Artificial Intelligence and Statistics},
  pages={11415--11434},
  year={2023},
  organization={PMLR}
}

@inproceedings{tailor2024learning,
  title={Learning to defer to a population: A meta-learning approach},
  author={Tailor, Dharmesh and
                  others},
  booktitle={International Conference on Artificial Intelligence and Statistics},
  pages={3475--3483},
  year={2024},
  organization={PMLR}
}

\end{document}